# Highlights

## M3BUNet: Mobile Mean Max UNet for Pancreas Segmentation on CT-Scans

Juwita Juwita,Ghulam Mubashar Hassan,Naveed Akhtar,Amitava Datta

- Proposed M3BUnet for dual phase pancreas segmentation.
- Introduced additional pre-processing steps, including external contour segmentation for coarse stage segmentation and wavelet decomposition for fine segmentation.
- Proposed Mean-Max block attention mechanism for improved pancreas segmentation.
- Experimental results on publicly available pancreas CT-scans datasets to establish the effectiveness of the proposed method.

# M3BUNet: Mobile Mean Max UNet for Pancreas Segmentation on CT-Scans

Juwita Juwita[a,c,*], Ghulam Mubashar Hassan[a], Naveed Akhtar[b,a] and Amitava Datta[a]

[a]Department of Computer Science and Software Engineering, The University of Western Australia, Australia
[b]School of Computing and Information Systems, The University of Melbourne, Australia
[c]Department of Informatics, Syiah Kuala University, Banda Aceh, Indonesia

ABSTRACT

Segmenting organs in CT scan images is a necessary process for multiple downstream medical image analysis tasks. Currently, manual CT scan segmentation by radiologists is prevalent, especially for organs like the pancreas, which requires a high level of domain expertise for reliable segmentation due to factors like small organ size, occlusion, and varying shapes. When resorting to automated pancreas segmentation, these factors translate to limited reliable labeled data to train effective segmentation models. Consequently, the performance of contemporary pancreas segmentation models is still not within acceptable ranges. To improve that, we propose M3BUNet, a fusion of MobileNet and U-Net neural networks, equipped with a novel Mean-Max (MM) attention that operates in two stages to gradually segment pancreas CT images from coarse to fine with mask guidance for object detection. This approach empowers the network to surpass segmentation performance achieved by similar network architectures and achieve results that are on par with complex state-of-the-art methods, all while maintaining a low parameter count. Additionally, we introduce external contour segmentation as a preprocessing step for the coarse stage to assist in the segmentation process through image standardization. For the fine segmentation stage, we found that applying a wavelet decomposition filter to create multi-input images enhances pancreas segmentation performance. We extensively evaluate our approach on the widely known NIH pancreas dataset and MSD pancreas dataset. Our approach demonstrates a considerable performance improvement, achieving an average Dice Similarity Coefficient (DSC) value of up to 89.53%±1.82 and an Intersection Over Union (IOU) score of up to 81.16±0.03% for the NIH pancreas dataset, and 88.60%±1.48 DSC and 79.90%±2.19 IOU for the MSD Pancreas dataset.

## 1. Introduction

Automating computed tomography (CT) segmentation is crucial to accelerate the pipeline of diagnosing organ abnormalities or monitoring organs at risk. Generally, CT image segmentation is performed manually by radiologists using tools like ITKSnap, which is time-consuming, laborious, and also prone to human error, especially for the parts of body with high variability, such as the pancreas [1]. Progress in the automation of pancreas segmentation from CT images still offers substantial room for improvement [2], [3]. This is primarily due to the facts that the pancreas is a relatively small organ that generally shares high similarity in texture and intensity values with its adjacent organs, it gets occluded, and also exhibits high variability in shape. Moreover, scanner movement during patient examinations often results in uneven or jagged edges around the pancreas, which further complicates the accurate delineation of the organ [4].

In general, data-driven feature learning has now become the primary technique for medical image segmentation. One of the popular neural network architecture used for this purpose is U-Net which is a Convolutional Neural Network (CNN) [5], [6], [7], [8], [9], [10]. It comprises of three main components: an encoder; which acts as a feature extractor for low-level features, a decoder; which is responsible for restoring the image resolution while merging features from various representation levels through the third component, known as skip connections. The key role of skip connections is to mitigate the information loss caused by down-sampling process in the network.

Whereas the above-noted three components work well in synergy, researchers have also highlighted limitations of the standard skip connections in achieving optimal model performance. As a result, Liu et al. [11] proposed a cross-domain fusion strategy, while [12] introduced fuzzy logic into the standard skip connections of U-Net. Another approach to address the dilution or loss of location-specific information in the network, particularly evident during the down-sampling phase, is to employ an attention mechanism [13, 14]. However, it is noteworthy that adding attention mechanism to a network is accompanied by a heavy computational overhead [15]. For example, incorporating attention into a model with 25.88M parameters, as in the case of [16], requires approximately an additional 190K parameters [17]. Other models have also reported a similar magnitude of computational overhead when integrating attention [18]. Recently, Dai et al. [19] leveraged a self-attention mechanism in their CNN-Transformer architecture, achieving state-of-the-art Dice Similarity Coefficient (DSC) scores for pancreas segmentation, albeit with a considerably larger number of model parameters, i.e., 86M.

*Corresponding author

juwita.juwita@research.uwa.edu.au, juwita@usk.ac.id (J. Juwita)
ORCID(s): 0000-0002-1118-3485 (J. Juwita)

A problem with models that rely on large size for improved performance is that their training also demands correspondingly large data and high computational resources. Their training time is often a bottleneck in practical resource-constrained settings, and they are susceptible to data overfitting, especially in real-life clinical scenarios. In this work, instead of resorting to the larger model size, we propose a lightweight model that is applied in a coarse-to-fine manner with mask guidance for effective segmentation. We devise a MobileNet [20]-UNet [5] framework as the foundational architecture. This architecture leverages a proposed mean-max (MM) block in the bottleneck layers, and it is applied in two stages for the segmentation task. The first stage results in a coarse predictive mask, which is exploited by the second stage. Owing to the underlying mechanism of our technique, we refer to our network as predictive Mask Mean-Max Block UNet (M3BUNet).

Our MM-block employs mean-max pooling operations across channels and subsequently fuses the feature maps from both operations with point-wise convolution results. The mean-pooling retains the more general features of the organ along with the spatial dimensions of the features. In parallel, max-pooling across channels accentuates the more prominent organ features. Both mean and max pooled features are able to encode complementary information. The block harnesses the benefits of both operations by eventually fusing the respective feature maps and using channel reduction.

Our model comprises of only 2.86M parameters, which is a significant reduction compared to the standard U-Net model with approximately 31M parameters [5] and its MobileNet variant [21] requiring around 7M parameters. Furthermore, our parameter count remains notably lower compared to other methods that rely heavily on attention mechanisms to preserve spatial information in the model [22], [23], [18]. Another distinctive contribution of this work is the introduction of the integration of an external contour segmentation sub-process for pre-processing in the first phase of the pancreas segmentation task. This process is employed to handle the large size variations of the region of interest in the inputs. By devising a relatively simple strategy, we demonstrate that the proposed contour segmentation effectively crops the relevant image regions, leading to performance improvements in the subsequent process. In the second stage of the our technique, we incorporate wavelet decomposition as an additional input to the network. This additional input enhances the edge features learned by the model and contributes to improved DSC scores.

Our key contributions can be described as following.

- We propose a novel neural network termed M3BUNet, which features a reduced parameter count while achieving comparable performance to SOTA methods for pancreas segmentation on popular benchmarks.

- We introduce external contour segmentation as a technique to mitigate input noise and suggest the use of multi-input based on two-dimensional wavelet decomposition to address blurred boundaries in pancreas segmentation.

- We devise mean-max (MM) block attention to enhance network performance for segmentation.

## 2. Related Work

Below, we organize the existing related literature for pancreas segmentation according to the key underlying techniques adopted by the methods. We mainly focus on closely related works to implicitly provide the background knowledge for our method.

### 2.1. Encoder-Decoder Techniques

In recent years, pancreas segmentation research has been dominated by adopting and refining CNN network architectures like U-Net [16]. The U-Net comprises of a contracting and an expanding path, with skip connections to enable the utilization of multilevel feature maps. Huang et al [21] conducted a study that integrated U-Net with MobileNet-V2 to significantly reduce the parameters of the model from 31M to 6.3M on the NIH Pancreas dataset [24]. The study achieved a DSC score of 82.87%. Another study with an aim to achieve a lightweight design was presented by Zhang et al. [25]. They attained a mean DSC score of 84.90% while using a 25.13M parameter model for the NIH Pancreas dataset. Recently, [26] tested their lightweight architecture design on the NIH Pancreas dataset and managed to reduce the training time to 154 minutes. However, this is still significantly longer than our proposed method, which requires approximately 20 minutes for its each phase. This time difference stems from fundamental technical differences between the two approaches.

### 2.2. Attention Mechanism

Existing research indicates that augmenting the U-Net architecture with attention mechanisms can lead to improved model performance [12], [27], [25]. Attention mechanisms empower a neural network to focus on crucial input features while disregarding irrelevant components [28], treating features as a dynamic selection process through adaptive weight assignment based on input significance. This mechanism originally found its roots in natural language processing (NLP), where an encoder-decoder attention module was developed to enhance neural translation engines [29], [30].

In their survey paper, Guo et al. [31] categorized attention into two main types: those based on transformer architectures and those based on CNN architectures. While self-attention transformers have been adopted in some cases for pancreas segmentation tasks, such as the approach proposed by Qiu et al. [32], who introduced the Residual-transformer UNet and achieved a DSC score of 86.25% for the NIH pancreas dataset, it did not outperform the CNN-based architecture proposed by Chen et al. [12]. The latter employs Global Average Pooling to emulate attention, achieving a noteworthy DSC score of 87.9 for the same dataset. On the other hand, Dai et al. [19] combined convolution attention-based methods with multi-head attention and achieved the state-of-the-art DSC score of 89.89%, albeit with a large number of parameters.

This demonstrates that, despite the presence of Vision Transformers with their global attention capabilities as one solution to address the limitations of locally operating attention convolutions, the performance improvement of attention transformers does not work in isolation but still requires suitable architectural settings. Therefore, despite the growing popularity of self-attention transformers, some researchers continue to utilize CNN-based attention mechanisms due to their potential learning power, ease of implementation, computational efficiency, and consistent results.

In this study, we propose attention-based convolutions by leveraging the mean-max operator. The effectiveness of max and mean pooling has also been reported in other contexts [33], [34], [35], [36]. It enhances crucial spatial features while suppressing less informative ones. This process enables the model to focus on significant regions within the input data, leading to improved performance in vision-related tasks.

### 2.3. Multi-Stage Segmentation

Many researchers advocate the 'coarse-to-fine' segmentation approach to tackle the task of medical image segmentation, especially for organs that are small in size and exhibit significant variation. Coarse- and fine-scaled segmentation models can be trained either separately or in conjunction [37]. In the separately trained approaches [38, 39, 40, 36], the coarse-scaled segmentation roughly localizes the organ to suppress the background. Subsequently, the fine-scaled segmentation utilizes the localized coordinates to crop the coarse-scaled input and conduct a more refined segmentation. In contrast to the separate training approaches, some researchers [41, 35, 42] have trained the coarse-scaled and fine-scaled segmentation models together. They utilize the coarse-scaled masks as weights to enhance the foreground of the fine-scaled input in an end-to-end manner.

While the separate training method is often considered more efficient, for eliminating the irrelevant background, it has a downside. That is, the fine-scaled segmentation can lack contextual information from the coarse-scaled segmentation during the training process. This issue is handled well by joint training, however the joint training does not remove the irrelevant background from the image, which can lead to other potential challenges like large computational time. Another drawback of the joint coarse-to-fine segmentation is the necessity to configure different experimental settings during the testing phase as compared to the training phase. This arises from the unavailability of masks during testing, requiring inter-iteration DSC throughout the testing process, which can sometimes introduce additional complexities. Our method is able to handle these issues well due to its unique two stage process.

## 3. Proposed Method

We present a novel technique for segmenting the pancreas in CT images, illustrated in Figure 1. Our approach comprises three key stages: pre-processing, segmentation phase one, and segmentation phase two. Each stage is discussed in detail in the sections to follow. Unique to our method is its peculiar two-stage process. The initial stage is used to perform coarse segmentation, while the subsequent stage handles fine-grained segmentation. Incidentally, both stages utilize the same architecture, which is a simple and elegant solution. Another distinctive feature of our technique is the incorporation of an external contour segmentation step during the pre-processing stage of the initial phase. Furthermore, in the second phase, we introduce wavelet decomposition deployment, an approach that has been rarely explored in previous studies on pancreas segmentation. Additionally, we introduce a mean-max (MM) block as an attention mechanism within our network architecture, termed M3BUNet. This mechanism enhances the model's ability to focus on pertinent information in CT images. Collectively, these innovative contributions make our technique highly effective while being efficient.

### 3.1. Pre-processing

Our pre-processing stage is inspired by several existing studies [12, 43, 44, 33]. We leverage established pre-processing techniques in medical image analysis, commonly known as Intensity Clipping, Voxel Spacing, Slicing, and .PNG Conversion. Upon implementing these steps, we performed a visual analysis of the images from the slicing process, identifying potential areas for improvement. One notable observation in our analysis was the variation in abdominal sizes across different patient images and the presence of abundant black pixels outside the abdominal shape and white pixels outside the abdominal wall, resembling steel-plate scans as visible in representative images in Figure 2. We consider these artefacts to be noise that could lead to inaccuracies in our modeling. Hence, to address this, we devise an external contour segmentation strategy.

#### 3.1.1. External Contour Segmentation

To determine the external contour, we first convert the image into a binary image by setting the threshold within the range [$\delta$, 255], where $\delta$ is empirically determined to be 77. The choice of this threshold is made to eliminate irrelevant background and white-steel-plate scans. We specifically extract the external contour while disregarding any internal contours within the objects as shown in Figure 3. It is found that threshold values significantly greater than $\delta = 77$ detect contours of other objects within the abdominal wall area, while threshold values much lower than this still leave many black pixels in the background outside the abdominal wall. Figure 3B shows that the resulting lines resembling circles are interconnected to the external contour boundary. We remove the irrelevant regions thus detected. The final outcome of our processing is displayed in Figure 3C. Thereafter, we standardize the image dimensions to $376 \times 376$. This results in images with coarser segmentation of the external abdominal contour. As demonstrated later in experiments, this simple strategy leads to a considerable improvement in the overall performance of our technique.

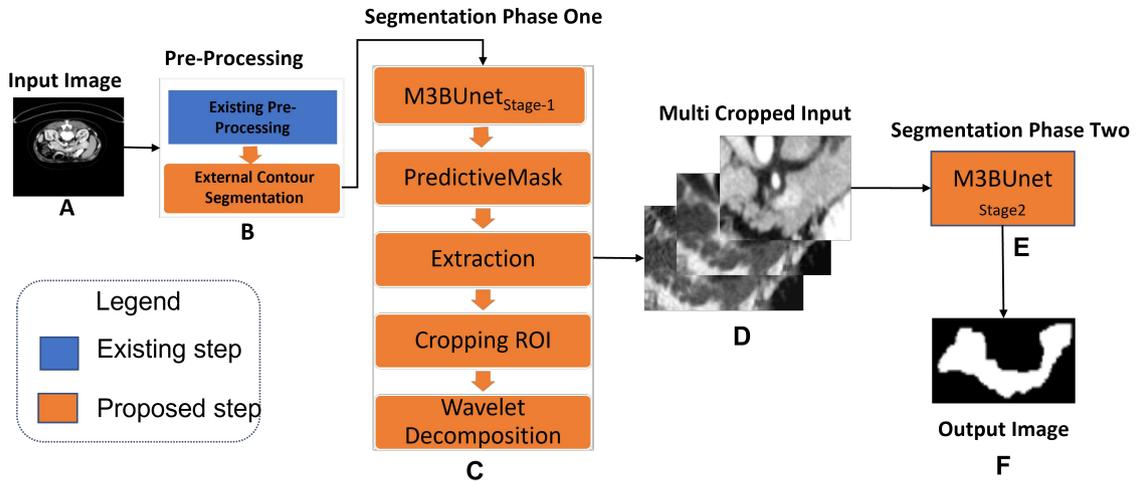

**Figure 1:** Overall WorkFlow Dual Stage Pancreas Segmentation Using M3BUNet: blue shapes depict conventional steps and orange blocks indicate the additional novel steps proposed in this work.

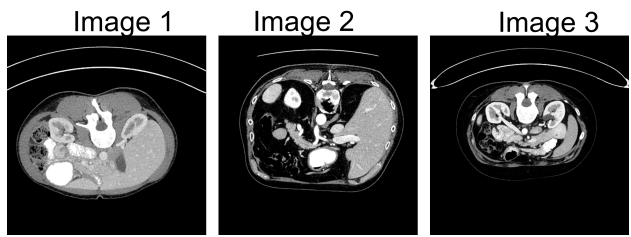

**Figure 2:** Representative original images after existing pre-processing step.

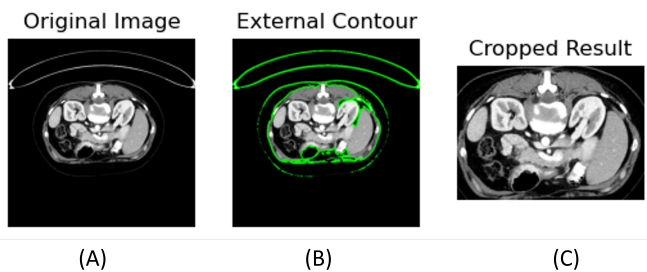

**Figure 3:** Sequence of images representing the steps in external contour segmentation.

### 3.2. Segmentation Phase One

Selecting an appropriate Region of Interest (ROI) is a critical endeavor for pancreas segmentation. A large ROI may incorporate unnecessary background to deteriorate model performance, while a smaller one can exclude essential segments of the pancreas. The primary objective of our phase one segmentation is to conservatively crop the ROI from the coarse crop image that is available to us from the pre-processing step, including the external contour segmentation step. In this phase, we train our model using M3BUnet$_{stage1}$. To keep the flow of discussion, we momentarily defer the discussion on M3BUNet to the next section. Here, we emphasize on the key idea that for this stage, the network is trained on the input data with a pancreatic segmentation objective. The network's output in this case is used to create a predictive mask, which we aim to use for the coarse localization of the pancreas.

Specifically, we extract the likely bounding box for the pancreas from the mask by considering the extreme coordinates of the predicted mask. This bounding box is expected to contain the pancreas because the network is trained for pancreas segmentation. We crop this bounding box region for further processing. During this detection phase, we address potential model limitations and the handling of scenarios where ROI detection might fail. We do this by implementing two conditional checks. Firstly, we incorporate a padding by adding 15 pixels in all coordinate directions to ensure better coverage. Secondly, in case the model fails to detect the ROI, we set a default ROI with dimensions of height (H): 168 and width (W): 229. These values were determined through a visual evaluation of the predicted pancreas location within the CT scan volumes. These values do not need modifications from sample to sample because the image size has already been standardized, as explained in the preceding section.

After obtaining the ROI for all images, we perform wavelet decomposition on the ROI to create additional input images that are ready to be fed into the M3BUNet. Details of our wavelet decomposition process are presented next.

#### 3.2.1. Wavelet Decomposition for Multi Input

While we successfully reduce a significant amount of background in the obtained cropping region results from the segmentation phase one, the boundary between the pancreas and adjacent organs still appears blurry. To address this, we employ Discrete Wavelet Transform Decomposition (DWT) [45] which is a technique used to analyze signals or images into low-frequency and high-frequency components. In the context of images, this allows for the decomposition of an image into a series of images with different spatial resolutions. In DWT, the original image is divided into several components, including the low-frequency component (usually referred to as LL, which stands for "Low-Low") and the high-frequency components (LH, HL, HH, which are referred to as "Low-High," "High-Low," and "High-High," respectively). The LL component represents the image with lower spatial resolution (or lower high-frequency content), while the LH, HL, and HH components represent high-frequency components that capture image details. In general, the formula for the scaling and translation functions of a wavelet for 1-D follows the scaling function as in equation 1 and wavelet function as in equation 2 :

$$W_\varphi(j_0, k) = \frac{1}{\sqrt{M}} \sum_x f(x)\varphi_{j_0,k}(x) \quad (1)$$

$$W_\psi(j, k) = \frac{1}{\sqrt{M}} \sum_k f(x)\psi_{j,k}(x) \quad (2)$$

where $W_\varphi(j_0, k)$ is the result of the transformation using the scaling function $\varphi$ at level $j_0$ and position $k$. $W_\psi(j, k)$ is the result of the transformation using the wavelet function $\psi$ at level $j$ and position $k$. $\frac{1}{\sqrt{M}}$ is the normalization factor. $f(x)$ is the signal that will be transforming. $\varphi_{j_0,k}(x)$ and $\psi_{j,k}(x)$ are the scaling and wavelet functions at level $j_0$ and $j$, respectively, and at position $k$.

From a 1-D setting, this DWT can be transformed into a 2-D form through a process as depicted in Figure 4 where $h_\psi$ is a low pass filter, $h_\varphi$ is high pass filter, $*$ is convolution operation and $2\downarrow$ is a down sample operation.

We perform two levels of wavelet decomposition using Daubechies filters [46], each with four coefficients. In this experiment, we only extract the Vertical Detailed Coefficient high-frequency levels 1 and 2 as additional inputs. Subsequently, we include these decomposed image results, along with the cropped original image resulting from the phase one segmentation, as inputs into our network. Therefore our additional inputs consist of high-frequency components of levels 1 and 2 decompositions, which primarily emphasize the edge texture in the image as shown in Figure 5. The image shows that the high-frequency decomposition of the original image provides clearer edge details. The higher the decomposition level, the more apparent the texture and boundaries of the pancreas.

### 3.3. Segmentation Phase Two

This stage comprises a direct application of the proposed M3BUNet. Hence, in this section, we discuss our network in detail. As shown in Figure 6, our network takes the form of an encoder-decoder backbone combined with MM-Blocks. In the figure, we also provide the architecture of the proposed

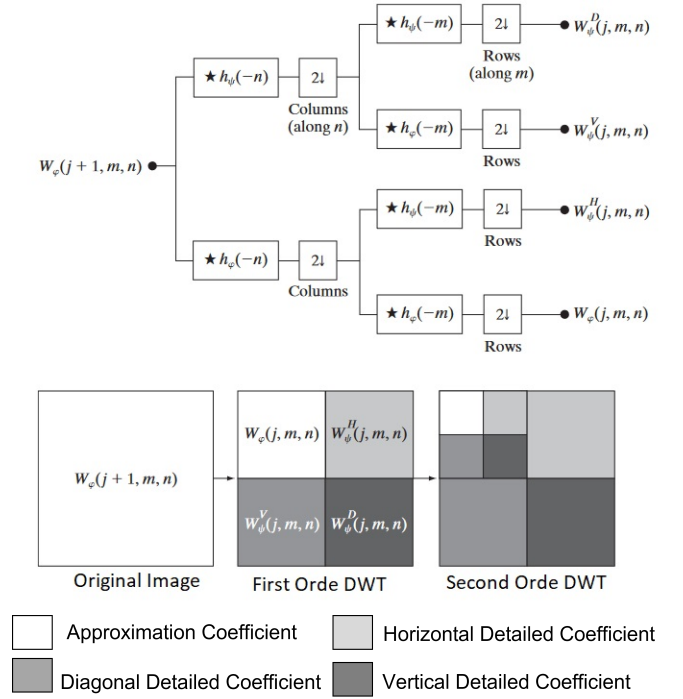

**Figure 4**: 2-D Wavelet Decomposition

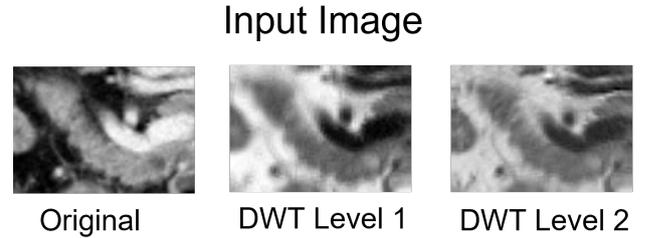

**Figure 5**: Representative multi-input images fed to M3BUNet

MM-Block. The components of the network are discussed in detail below.

#### 3.3.1. M3BUNet Encoder-Decoder

**Encoder:** Given the limited availability of medical image training data, we technically employed transfer learning. We utilize the MobileV2Net model [47] pre-trained on another dataset (ImageNet) as our foundation for the task. Leveraging weights from pre-trained models on large non-medical datasets has been demonstrated to improve the performance of image segmentation tasks [48]. Subsequently, we carry out fine-tuning. From the 17 blocks available in the original architecture, we select blocks 0 through 13. At the Encoder's initial layer, the number of channels is set to 32. While we do not make significant changes to the original structure by [47], our parameter reduction occurs in the encoder due to the presence of inverted residual blocks inherited from the original MobileNetV2 structure. A unique aspect of our architecture design is the combination of a streamlined arrangement in the encoder section, which employs depthwise convolution, with a slightly heavier arrangement in the decoder section, still using standard 2D convolution. An excessively small model may compromise accuracy, thus a balance between the number of parameters and accuracy is considered in the encoder.

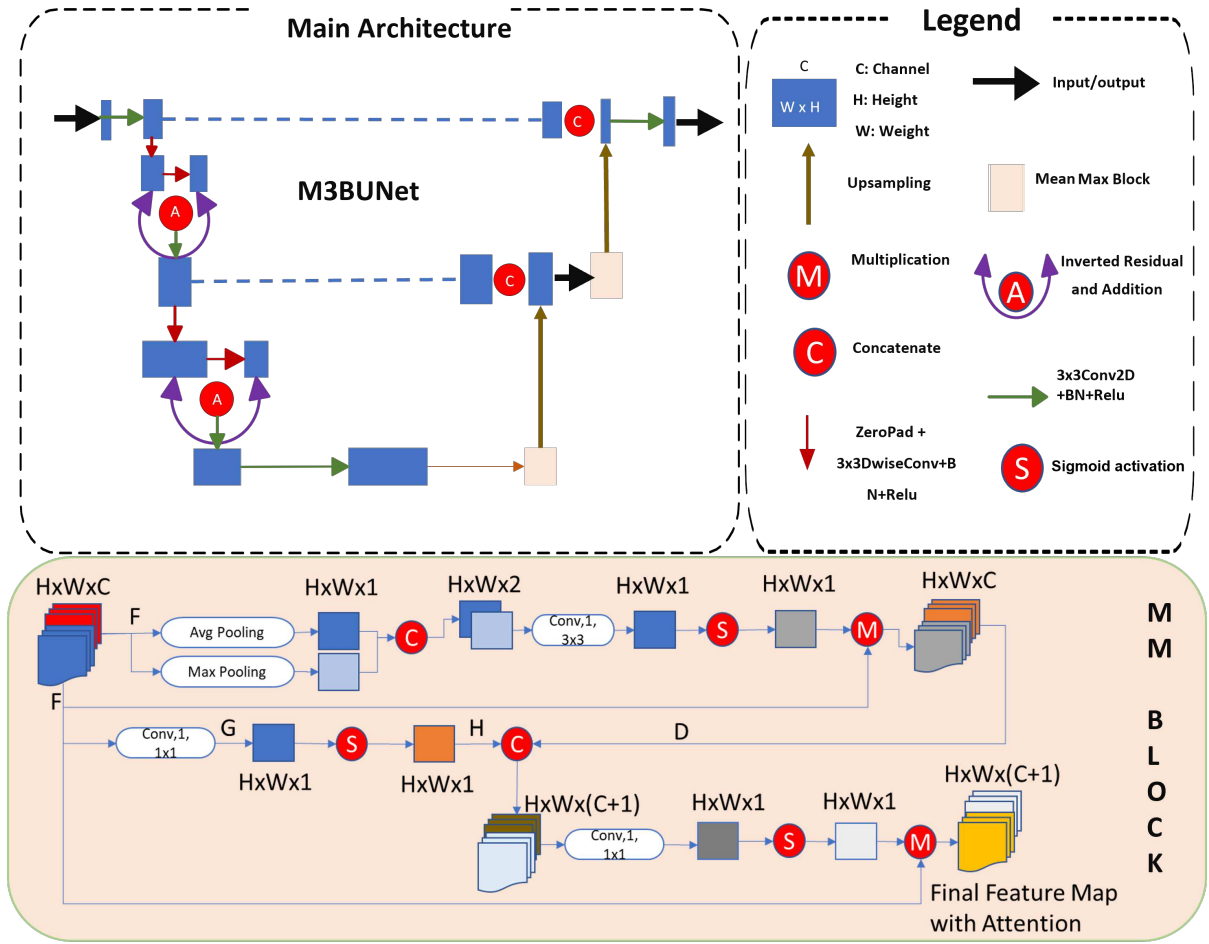

**Figure 6:** Proposed M3BUNet with MM-Block

**Decoder:** The decoder section of our model follows the U-Net architecture [5] with all convolutional filters having a size of 3x3. To maintain a reasonable model size, we chose a moderate number of filters, namely 128, 64, 32, and 16. In the decoder section, we incorporate the proposed novel MMBlock. The structure of our MMBlock is elaborated in the following subsection.

### 3.3.2. MM BLock

It is a convolution-based attention mechanism consisting of two attention pathways. The first pathway is based on average and maximum pooling across all channels. We choose a 3x3 convolution instead of a larger kernel size to avoid losing neighboring pixel details, which differs from the approach used in [18]. Initially, the input feature, resulting from the crop-concatenation of the encoder and decoder, enters the MM-Block. This is denoted as 'F'. Subsequently,

'F' undergoes some operations. One of them is average pooling across all channels, and the second is max pooling. The results of these two operations are combined to enhance the learned descriptor features. Equations 3 and 4 given below are used to calculate mean and average pooling along the channel axis. These two operators, though functioning differently, complement each other. Mean pooling across channels aims to capture the global features of pancreas. In contrast, max pooling across channels is designed to capture the most prominent features of the organ, as shown in Figure 7. As evident from the figure, before undergoing the mean-max pooling operation, the activation feature map still contains a significant amount of background information. Subsequently, the mean pooling operation provides a general overview of the pancreas's shape, while the max pooling operation highlights the edges of the pancreas more clearly. The fusion of feature maps from both operations directs attention to the crucial features of the pancreas.

$$\text{MeanPooling}(x) = \frac{1}{C} \sum_{c=1}^{C} x(i, j, c), \qquad (3)$$

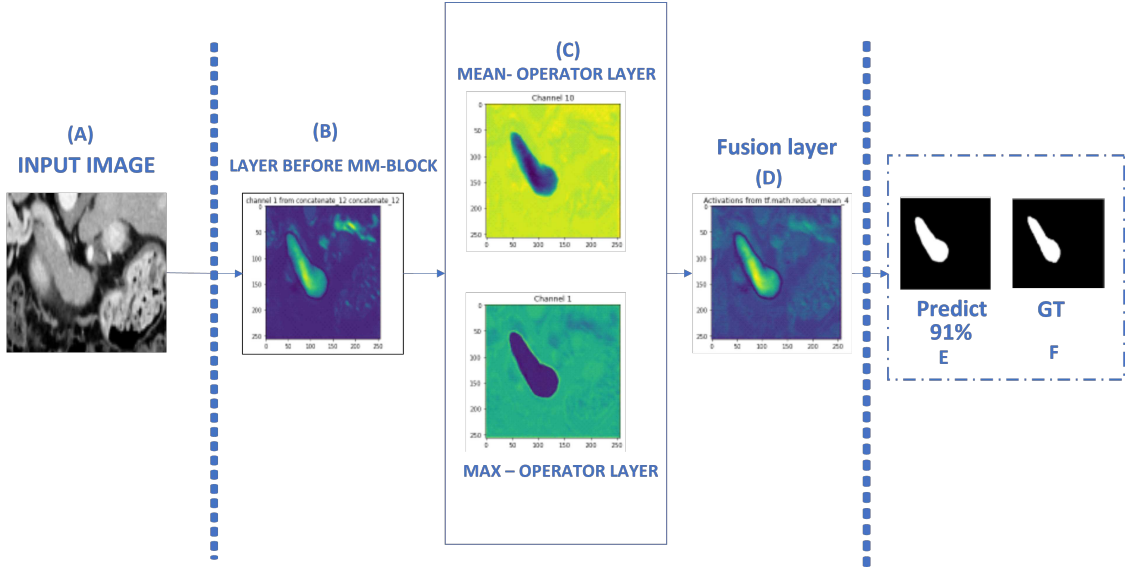

**Figure 7:** The proposed Mean-Max block extracts and fuses complementary features that preserve locational information. (A) Input Image selected randomly from case 73, slice 115 for an axial view in NIH Pancreas dataset. (B) Representative feature channels. (C) Results of mean and max pooling. (D) Fusion. (E-F) Prediction with 91% accuracy along with ground truth

$$\text{MaxPooling}(x) = \max_{c=1}^{C} x(i, j, c), \quad (4)$$

$$\text{ConcatenatedResult}(i, j) = \begin{bmatrix} \text{MeanPooling}(x)(i,j) \\ \text{MaxPooling}(x)(i,j) \end{bmatrix} \quad (5)$$

where $i, j$ denote the feature spatial indices and $c$ is the channel. Given $x'$ as an input feature map and $W$ as a 3x3 convolution filter weights, the used convolutional operations can be described as

$$\text{ConvRes}(i, j) = \sum_{m=-1}^{1} \sum_{n=-1}^{1} W(m, n) \cdot x'(i+m, j+n) \quad (6)$$

The computation of point-wise convolution can be expressed as:

$$F(x) = \sigma(\text{ConvRes}^{1x1}) + x \quad (7)$$

where $\sigma$ is the activation function.

The next attention pathway is the novel point-wise convolution pathway. where we apply pointwise convolution with a filter, followed by sigmoid activation. In this pathway, there is a reduction in channel dimensions, and then attention computation is performed using sigmoid activation. Furthermore, the feature map from the first pathway, denoted as 'D' in Figure 6 , is fused with the point-wise attention from the second pathway, denoted as 'H', and channel conversion is performed before being merged back with the original feature map. This strategy facilitates the network's understanding of interactions among varying channel features. The final result is a final feature map that has been attended to important regions of the pancreas.

## 4. Experiments

In this section, we discuss the datasets to validate our proposed network, followed by evaluation metrics and implementation.

### 4.1. Datasets

We employ two publicly available datasets for evaluation. Details of the datasets and any specific pre-processing operations are given below.

**NIH Healty-Pancreas Dataset:** This dataset comprises of 82 abdominal contrast-enhanced CT scans in DICOM format. Trained radiologists provide ground truth annotations for this dataset, available in NIFTI format. Two CT scans were excluded from the dataset as they lacked mask images. The original size of the serial CT scans varied between 181 and 466 slices for different patients, with each slice containing 512 x 512 pixels. Initially, the DICOM and NIFTI Volume Files were sliced and converted to 512 x 512 PNG images. Due to the imbalance between pancreas and background, we only retained sliced axial view images containing > 5% pancreas. This is inline with the standard practice [12], and it results in a total of 1700 images. Next, we clipped the image intensities within the range [-100, 240] and set the voxel spacing to (1, 1, 1). Then, we applied the external contour segmentation.

**MSD Tumors-Pancreas Dataset:** This dataset comprises of 281 abdominal contrast-enhanced CT scans that include labeled pancreas and pancreatic tumors. This dataset is sourced from the Medical Segmentation Decathlon (MSD) pancreas segmentation dataset [49]. Each CT volume has a resolution of 512 x 512 pixels, and the number of slices in the CT scans ranges from 37 to 751. We filtered the dataset to retain only axial view images with more than 5% pancreas content. In line with prior research [50], [19], we amalgamated the pancreas and pancreatic tumor mask into a single entity for segmentation purposes. The remaining processing steps align with those used for the NIH Pancreas dataset.

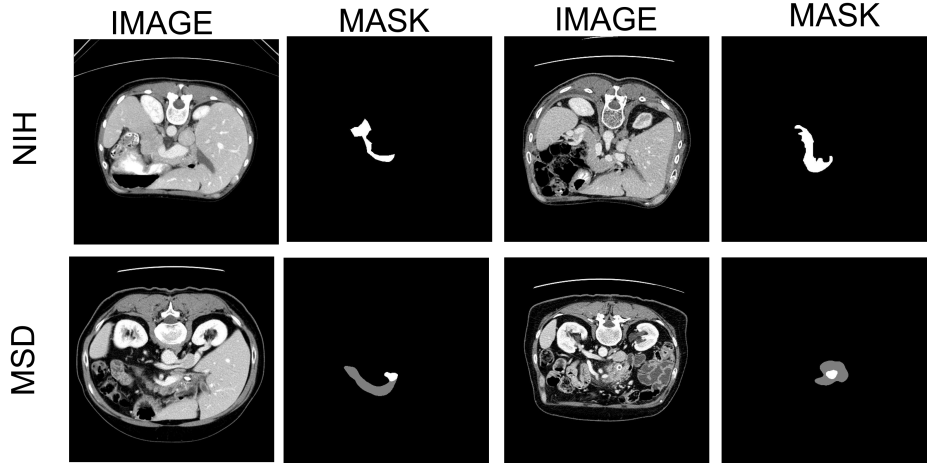

**Figure 8:** The 2D visualization includes examples of images and masks for both the NIH Pancreas and MSD Pancreas datasets. The upper row contains samples representing images and masks for the NIH Pancreas dataset, while the bottom row contains sample images and masks from the MSD Pancreas dataset. Masks that appear brighter or whiter indicate the presence of tumors.

A visualization of representative images and masks for the two used datasets, which have undergone pre-processing, is shown in Figure 8.

### 4.2. Evaluation Metrics

To assess the performance of models, we employ five standard quantitative metrics. These metrics include Dice-Sorensen Coefficient (DSC), which gauges the similarity between segmentation predictions and the ground truth; and Intersection Over Union (IOU), which measures the ratio of intersection and union of predicted foreground pixels and ground truth foreground pixels. We also evaluate the Specificity to determine the accurate identification rate of both non-pancreas and pancreas pixels. Moreover, Precision is used to gauge the proportion of genuinely positive pixels in our predictions, while Recall is used to quantify the fraction of correctly identified pancreas pixels against the total pancreas pixels present in the reference image.

### 4.3. Implementation

In our experiments, the hyperparameter settings for both segmentation phases is kept consistent. We conducted experiments using TensorFlow open-source library on an NVIDIA Titan V GPU with 12GB memory. We employed Adam optimizer and a batch size of 8.

Our training process consisted of two phases. In the initial 10 epochs, we performed transfer learning with all trainable layers frozen, using a learning rate of 0.001. Subsequently, in the following 100 epochs, we conducted fine-tuning with the learning rate starting at 0.0001. To ensure effective monitoring of the training process, we employed several callbacks, including saving the best-performing model based on the reduction of training loss, dynamically adjusting the learning rate by a factor of 0.1, down to a minimum threshold of $1 \times 10^{-7}$, and implementing early stopping if there was no improvement for 15 consecutive epochs.

Next, to prevent overfitting, we implemented data augmentation techniques, including vertical flip, horizontal flip, and random 90-degree rotation. These augmentation techniques were applied to 50% of the original training images.

In the initial segmentation phase, we employed plain-MBUNet (without MMblock), whereas in the second segmentation phase, we utilized M3BUNet. The input images for phase one were images that had undergone external contour segmentation and had been resized to 256 x 256 pixels. Meanwhile, the input images for phase two were images cropped from the first phase and resized into 64 x 64 pixels, adding two images resulting from wavelet decomposition.

For each phase, we conducted a 4-fold cross-validation based on patient data. The division of the training and testing sets was performed randomly, considering patient cases, resulting in four subsets, each containing approximately 20 patients for the NIH Pancreas dataset, and four subsets with approximately 70, 70, 70, and 71 patients for the MSD Pancreas dataset.

## 5. Results and Discussion

This section presents the experimental results and discussion for both datasets and ablation study.

### 5.1. Intermediate Result and Ablation

Our proposed model consists of two stages, therefore we presents and discusses the intermediate results and its ablation first.

#### 5.1.1. Intermediate Result

In Table 1, we compare the performance of our segmentation results for the first stage with the same stage of the other state-of-the-art two-stage techniques. In the table, the training strategies for the models are categorized as 'jointly' and 'separately'. The former uses joint training for both the models in the two-stage techniques, whereas the latter trains the models separately, similar to our technique. For the ease of understanding, we only display the segmentation accuracy results for the first stage for the NIH Pancreas dataset. The same procedure was also applied to the MSD Pancreas dataset.

**Table 1**
Segmentation results comparison for phase one on the NIH Pancreas Dataset. Here, '4-CV-PI' denotes 4-Fold Cross Validation Patient Independent protocol, '-' indicates information not available, and 'PM' means Predictive Mask.

| Method | Validation Scheme | Training Strategy | Region of Interest (ROI) | Mean DSC ± Std (%) |
|---|---|---|---|---|
| DCNN, 2021 [25] | 4-CV-PI | Separately | Define by spatial prior | 72.69 |
| Fixed-Point, 2017 [52] | 4-CV - | Separately | Defined by ground truth | 73.08±9.60 |
| 2.5D U-Net, 2021 [51] | 4-CV-PI | Jointly | PM + frame | 76.89±7.44 |
| MLN, 2022 [11] | 4-CV-PI | Jointly/Separately | PM | 77.96 |
| U-Net, 2023 [32] | 4-CV-PI | Separately | PM | 78.09±5.69 |
| MS Cues, 2018 [53] | 4-CV-PI | Jointly | PM + 20 Padding | 78.23 |
| M3BUNet-Ours | 4-CV-PI | Separately | PM + 15 Padding | **78.97±2.68** |

The results in Table 1 demonstrates that in the initial phase, by employing the external contour segmentation model, we achieved a DSC percentage of 78.97%, which represents the best overall performance for this segmentation phase. When compared to the baseline U-Net, which is still widely used in other studies as either a coarse cropping method [51] or for localization, our model outperforms it by 2.08% and 0.88%.

### 5.1.2. Ablation

Our approach employs External Contour Segmentation. To assess its impact on segmentation accuracy, we conducted ablation experiments following patient-independent protocols for both datasets. Table 2 demonstrates that including External Contour Segmentation as part of the common pre-processing step led to an improved pancreas segmentation accuracy of 5.62% for the NIH pancreas dataset, increasing from 73.35% to 78.97%. Similarly, when we conducted an ablation experiment on the MSD Pancreas dataset without this component, the DSC was only 49.04%. A substantial improvement of 22% was observed with the application of external contour segmentation.

These results clearly indicates that accuracy improves for both datasets when our external contour segmentation is employed. We discuss the ablation experiment for the MM-Block in Section 5.2.2.

### 5.2. Final Segmentation and Ablation

Here, we presents and discusses the overall results of our proposed model and its ablation.

### 5.2.1. NIH Pancreas Dataset

We evaluated the performance of our model on the NIH pancreas dataset and present its detailed comparison with the existing methods in Table 3. In the reported results, we have carefully selected the existing state-of-the-art methods that are optimized for the 2D (i.e., image) treatment of the pancreas segmentation task for a fair comparison.

**Table 2**
Comparison of performance of Contour Segmentation and Non Contour Segmentation for stage 1 with M3BUNet model.

| Dataset | Contour Segmentation | Mean DSC (%) | IOU (%) | Specificity (%) |
|---|---|---|---|---|
| **NIH** | Yes | 78.97 | 68.17 | 96.61 |
|  | No | 73.35 | 62.20 | 99.79 |
| **MSD** | Yes | 71.04 | 57.35 | 99.44 |
|  | No | 49.04 | 28.64 | 99.38 |

Our model proves to be highly effective in pancreas segmentation. Additionally, our model demonstrates exceptional pixel-level pancreas identification capabilities, as evident from our recall rate of 92.80%. Notably, our model stands out for its simplicity of implementation and significantly lower parameter count compared to state-of-the-art models. A comparison of model parameters is provided later in Table 5, where we discuss model sizes in more details.

### 5.2.2. MSD Pancreas Dataset

Table 4 presents the comparison of the proposed network with current state-of-the-art methods. Our DSC results rank as the second best, whereas our DSC performance is slightly lower than the best performing method. Our parameter count is significantly lower, which is approximately 30 times less than the model providing state-of-the-art results.

### 5.2.3. Model Size Comparison

Our model is designed to be more parameter-efficient than other SOTA counterparts. Although parameter count considerations in recent studies may be scarce, we offer comparisons with models that have disclosed their parameter counts in Table 5. At 2.86M parameters, ours proposed model stands out as the most computationally efficient. It is important to mention that we report the parameter count for our method by adding the parameters for both stages.

Our model is exceptionally compact, especially when compared to the state-of-the-art model [19], which has an enormous parameter count of 86M. In addition, our model not only outperforms a similar basic structure [21] but does so with significantly fewer resources. Huang et al. [21] applied 12 types of augmentation on the NIH Pancreas dataset, while our model achieves a 6.6% higher DSC using only 3 types of augmentation. Remarkably, our proposed model operates with a dataset approximately 10 times smaller than theirs.

Table 3
Comparison of techniques on the NIH Pancreas dataset. The results are based on four-fold cross-validation under patient-independent protocol. Similar to the proposed technique, reported existing methods are optimized for 2D image-based pancreas segmentation. Results of the existing methods are taken directly from original papers, following the same evaluation protocols. The best results are marked in bold.

| Network | Basic Structure | Mean ± Std | | | | |
| --- | --- | --- | --- | --- | --- | --- |
| | | DSC | IOU | Specificity | Precision | Recall |
| MBU-Net, 2022 [21] | CNN- MobileNet | 82.87±1.00 | 70.97±1.39 | **99.95±0.01** | 89.29±0.98 | 77.37±1.41 |
| DTUNet, 2022 [54] | CNN-Transformer | 84.77±4.65 | N/A | N/A | N/A | N/A |
| Spatial PriorNet, 2021 [25] | CNN | 84.90 | NA | NA | NA | NA |
| VGG-UNet, 2022 [40] | CNN-VGG-Net | 85.40±1.6 | 73.21±6.8 | N/A | N/A | N/A |
| DenseASPP, 2021 [55] | CNN | 85.49±4.77 | N/A | N/A | N/A | N/A |
| LocNet+ ECTN, 2023 [43] | CNN | 85.58±3.98 | 74.99±5.86 | N/A | 86.59±6.14 | 85.11±5.96 |
| MAD-UNet, 2021 [56] | CNN | 86.10±3.52 | 75.55±5.42 | 84.97±6.18 | N/A | N/A |
| RTUNet, 2023 [32] | CNN-Transformer | 86.25±4.52 | N/A | N/A | N/A | N/A |
| Multi Scale U-Net, 2021 [51] | CNN | 86.61±3.47 | N/A | N/A | N/A | N/A |
| tU-Net+, 2022 [12] | CNN | 87.91±2.65 | 78.52±4.14 | N/A | 90.43±3.77 | 85.77±4.61 |
| TD-Net, 2023 [19] | CNN-Transformer | **89.89±1.82** | N/A | N/A | **89.59±1.75** | 91.13±1.48 |
| M3BUNet (Ours) | CNN-MobileNet | 89.53±0.02 | **81.16±0.03** | 91.58±0.01 | 86.75±0.01 | **92.80±0.01** |

Table 4
Comparison of techniques on the MSD Pancreas dataset. The results are based on four-fold cross-validation under patient-independent protocol. Similar to the proposed technique, reported existing methods are optimized for 2D image-based pancreas segmentation. Results of the existing methods are taken directly from original papers, following the same evaluation protocols. The best results are marked in bold.

| Network | Basic Structure | Mean ± Std | | | | |
| --- | --- | --- | --- | --- | --- | --- |
| | | DSC | IOU | Specificity | Precision | Recall |
| MAD-UNet, 2021 [56] | CNN | 88.52±3.77 | 79.42±5.82 | 89.66±4.62 | N/A | N/A |
| Recurrent U-Net, 2021 [57] | CNN | 85.65 | N/A | N/A | N/A | N/A |
| tU-Net+, 2022 [12] | CNN | 66.48 | N/A | N/A | N/A | N/A |
| RTUNet, 2023 [32] | CNN-Transformer | 86.25±4.52 | N/A | N/A | N/A | N/A |
| TD-Net, 2023 [19] | CNN-Transformer | **91.22±1.37** | N/A | N/A | **93.22±2.79** | **91.35±1.63** |
| M3BUNet (Ours) | CNN-MobileNet | 88.60±1.48 | **79.9±2.19** | **95.71±0.78** | 90.76±1.42 | 87.20±2.83 |

This substantial improvement can be attributed to our adept utilization of transfer learning and fine-tuning techniques, which have consistently demonstrated their effectiveness in the domain of pancreas segmentation.

### 5.2.4. Comparison MM-Block with other Attention

We also conducted a comparison of MM-Block with several popular convolution-based attention mechanisms commonly used by other researchers. The attention blocks we compared were: Squeeze and Excitation Attention (SE) [23], Convolutional Block Attention Module (CBAM) [18], and Global Local Attention Module (GLAM) [60]. The comparison results are presented in Table 6. As a baseline, we employed M3BUNet without MM-Block and wavelet decomposition. To ensure a fair comparison, all hyperparameter settings and attention block placements were kept the same, namely, the bottleneck and decoder. From the table, it can be observed that the MM-Block attention outperforms the other three popular attention blocks.

### 5.2.5. Ablation

In our finer segmentation method, we utilize multi-inputs based on wavelet decomposition and the MM-Block. To assess the influence of these components on segmentation accuracy, we conducted experiments on both datasets. Additionally, we performed ablation experiments on the proposed components and compared them to the baseline results.

Table 7 demonstrates that the proposed M3BUNet architecture design successfully improves the DSC compared to the baseline MobileNet-UNet by 6.18% from 83.35% to 89.53% for the NIH Pancreas dataset and by 4.15% from 84.48% to 88.60% for the MSD Pancreas dataset. Removing the MM-Block from M3BUNet results in a decrease of DSC by 3.28% for the MSD Pancreas dataset and by 5.21% compared to the baseline (MobileNet-UNet). Similarly, wavelet decomposition significantly contributes to the enhancement of DSC for both datasets, with the use of multi-input decomposition providing an increase of 1.28% for the MSD Pancreas dataset and 3.69% for the NIH Pancreas dataset.

Table 5
Comparison of models' parameters. The best results are marked in bold.

| Algorithm | Mean DSC (%) | Parameters |
|---|---|---|
| MLConvNets [58] | 71.42 | 57.49 M |
| SAHNNs [38] | 71.42 | 42.10 M |
| Attention-UNet [59] | 81.48 | 33.26 M |
| C2F [52] | 82.50 | 256.11M |
| MBU-Net [21] | 82.87 | 6.30 M |
| MDAG-Net [22] | 83.04 | 7.9 M |
| RSTN [53] | 84.50 | 256.11 M |
| DCNN [25] | 84.90 | 25.13 M |
| DenseASPP [55] | 85.49 | - |
| LocNet+ECTN [43] | 85.58 | - |
| tai MAD-UNet [56] | 86.10 | - |
| MScale-UNet [51] | 86.61 | - |
| tU-Net [12] | 87.91 | - |
| TD-Net [19] | 89.89 | 86M |
| **M3BUNet-Ours** | **89.52** | **2.86 M** |

Table 6
Comparison of MM-block with other popular attention block. The experiment conducted on the NIH Pancreas dataset. The best results are marked in bold.

| Network | Mean DSC |
|---|---|
| M3BUNet | 87.04 ± 1.18 |
| M3BUNet + SE | 86.81 ± 0.95 |
| M3BUNet + CBAM | 87.41 ± 0.69 |
| M3BUNet + GLAM | 85.15 ± 0.62 |
| M3BUNet + MM-block | 88.56 ± 0.03 |

Table 7
Comparison of U-Net Variant on the NIH Pancreas dataset. The segmentation results are obtained by four-fold cross validation. The best result are marked in bold.

| Dataset | Model | Wavelet | MM-Block | DSC |
|---|---|---|---|---|
| NIH | UNet | - | - | 83.59± 0.01 |
| | MobileNet-UNet | - | - | 83.35 ± 0.03 |
| | M3BUNet(ours) | ✓ | - | 87.04± 1.18 |
| | M3BUNet(ours) | - | ✓ | 88.56 ± 0.03 |
| | M3BUNet(ours) | ✓ | ✓ | 89.53 ±0.02 |
| MSD | UNet | - | - | 84.45± 1.79 |
| | MobileNet-UNet | - | - | 84.48± 2.14 |
| | M3BUNet(ours) | ✓ | - | 85.76± 2.91 |
| | M3BUNet(ours) | - | ✓ | 87.76± 2.13 |
| | M3BUNet(ours) | ✓ | ✓ | 88.60 ± 1.48 |

## 6. Conclusion

In this paper, we introduced M3BUNet, an efficient neural model for pancreatic segmentation. Our network incorporates a unique Mean-Max (MM) block, which efficiently focuses on critical regions of input images for segmentation. Our approach consists of two stages. Additionally, we have employed two preprocessing techniques: external contour segmentation and wavelet decomposition. Both of these techniques have been shown to enhance the accuracy of pancreatic segmentation for using two datasets, namely the NIH Pancreas dataset and the MSD Pancreas dataset. Overall, our proposed M3BUNet outperforms existing models computationally while providing the comparable or better results.

## Consent for publication

All authors have read and approved the manuscript

## CRediT authorship contribution statement

**Juwita Juwita:** Conceptualization, Investigation, Data curation, Visualization, Software, Methodology Development, Writing– original draft, Writing–review and editing, Validation. **Ghulam Mubashar Hassan:** Conceptualization, Writing-review and editing, Supervision. **Naveed Akhtar:** Conceptualization, Review-original draft, Writing-review and editing, Resource, Supervision. **Amitava Datta:** Conceptualization, Supervision.

## Declaration of competing interest

The authors declare that they have no known competing financial interests or personal relationships that could have appeared to influence the work reported in this paper.

## Data availability

All datasets are publicly available. The code will be made available at github at the time of publishing the study.